\documentclass[pre,superscriptaddress,twocolumn]{revtex4-1}

\usepackage{graphicx} 
\usepackage{epsfig} 
\usepackage{amssymb}
\usepackage{amsmath}

\usepackage{color}
\newcommand{\cm}[1]{{\color{black} #1}}

\usepackage{pifont} 

\newcommand{\dawsonmax}{F_\xam^D}
\newcommand{\dPdTrho}{\left(\frac{\partial P}{\partial T}\right)_{\!\!\!\rho_0}}  
\newcommand{\deltaTnpnoloss}{\Delta \Tnp^*}
\newcommand{\ea}{e_{\rm a}}
\newcommand{\Ea}{E_{\rm a}}
\newcommand{\Ef}{E_{\rm f}}
\newcommand{\EL}{E_{\rm L}}
\newcommand{\Esigma}{E_\gamma}
\newcommand{\ga}{{\gamma}}
\newcommand{\Fth} {F_\mathrm{th}}
\newcommand{\nth}{n^{\rm th}}
\newcommand{\Pmax}{P_{\rm max}}
\newcommand{\Pprime}{P'}
\newcommand{\tauep}{\tau_\mathrm{ep}}
\newcommand{\tauk}{\tau_\mathrm{K}}
\newcommand{\tvap}{t_\mathrm{vap}}
\newcommand{\taup}{\tau_\mathrm{p}}
\newcommand{\taur}{\tau_\mathrm{r}}
\newcommand{\Tmax}{T_\xam}
\newcommand{\To}{T_0}
\newcommand{\Ts}{T_\mathrm{s}}
\newcommand{\Tsp}{T_\mathrm{sp}}
\newcommand{\xam}{\mathrm{max}}

\newcommand{\cf} {C}
\newcommand{\lamf} {k}

\newcommand{\cnp}{C_\mathrm{np}}
\newcommand{\Rnp}{R_\mathrm{np}}
\newcommand{\signp}{\sigma_\mathrm{abs}}
\newcommand{\Snp}{S_\mathrm{np}}
\newcommand{\Tnp}{T_\mathrm{np}}
\newcommand{\Vnp}{V_\mathrm{np}}

\newcommand{\vcr}{\mathbf{r}}
\newcommand{\vv}{\mathbf{v}}

\newcommand{\flap}{\mathcal{F}}

\newcommand{\fnp } {f_\mathrm{np}}
\newcommand{\fmeds}{f_\mathrm{s}}
\newcommand{\fmed }{f}

\newcommand{\ftilde}  {\tilde{f}}
\newcommand{\fnptilde}{\ftilde_\mathrm{np}}
\newcommand{\fmedstilde}{\ftilde_\mathrm{s}}

\begin{document}

\title{Strong and fast rising pressure waves emitted by plasmonic vapor nanobubbles}
\author{Julien Lombard}
\affiliation{Departamento de F\'{\i}sica y Qu\'{\i}mica Te\'orica, Facultad de Qu\'{\i}mica \\ 
Universidad Nacional Aut\'{o}noma de M\'{e}xico, Mexico City 04510, Mexico}
\author{Julien Lam}
\affiliation{CEMES, CNRS and Université de Toulouse, 29 rue Jeanne Marvig, 31055 Toulouse Cedex, France}
\author{Fran\c{c}ois Detcheverry}
\affiliation{Univ. Lyon, Universit\'e Claude Bernard Lyon 1, CNRS,\\ 
Institut Lumi\`ere Mati\`ere, F-69622, Villeurbanne, France}
\author{Thierry Biben}
\affiliation{Univ. Lyon, Universit\'e Claude Bernard Lyon 1, CNRS,\\ 
Institut Lumi\`ere Mati\`ere, F-69622, Villeurbanne, France}
 \author{Samy Merabia\footnote{Author to whom correspondence should
 be addressed: \mbox{samy.merabia@univ-lyon1.fr}}}
 \affiliation{Univ. Lyon, Universit\'e Claude Bernard Lyon 1, CNRS,\\ 
 Institut Lumi\`ere Mati\`ere, F-69622, Villeurbanne, France}

\date{\today}

\begin{abstract}
Plasmonic vapour nanobubbles are currently considered for a wide variety of applications 
ranging from solar energy harvesting and photoacoustic imaging to nanoparticle-assisted cancer therapy.
Yet,  due their small size and unstable nature,  
their generation and consequences remain difficult to characterize. 
Here, building on a phase-field model, 
we report on the existence of strong pressure waves 
that are emitted when vapor nanobubbles first form around a laser-heated nanoparticle immersed in water,
and subsequently after bubble rebound.
These effects are strongest when the fluid is locally brought high in its supercritical state, which may be realized with a short laser pulse. 
Because of the highly out-of-equilibrium nature of nanobubble generation, 
the waves combine a high pressure peak with a fast pressure rising time, 
and propagate in water over micron distances,  
opening the way to induce spatially and temporally localized damage. 
Discussing the consequences on biological cell membranes, 
we conclude that acoustic-mediated perforation is more efficient than nanobubble expansion  to breach membrane. 
Our findings should serve as guide 
for optimizing the thermoacoustic conversion efficiency of plasmonic vapor nanobubbles.  
\vspace*{1cm} 

\end{abstract}

\maketitle

\section{Introduction}

Due to their wide range of applications in biomedecine, surface cleaning and photoacoustic imaging~\cite{Boulais2013,Baffou2020}, 
plasmonic vapor nanobubbles have received an ever increasing attention from the scientific community. 
Their formation involves both
the remarkable optical properties of metallic nanoparticles 
and the unique thermomechanical response of nanoscale vapor blankets. 
Nanoparticles offer not only versatility in optical absorption, controlled by size, composition and laser wavelength, 
but also biocompatibility and possible conjugation with antibodies for specific targetting~\cite{Pitsillides2003}.      

When irradiated by strong laser pulses, metallic nanoparticles may be heated up by hundredths of Kelvin~\cite{Baffou2018}. 
At the basis of this phenomenon is  the interaction between the laser and the nanoparticle conduction electrons, 
and its exaltation when the incident electromagnetic wavelength is close to the nanoparticle surface plasmon resonance,
which lies in the visible range for common nanostructures~\cite{Baffou2012}. 
Nanoparticles in  solution   
thus behave as nanoscale hot spots for the liquid environment. 
In this unique situation, 
water may undergo locally quick phase change, as its local temperature approaches the spinodal, around $550\,$K for water~\cite{Lombard2014,Metwally2015}. 
Indeed, at such small scales, the energy barriers for heterogeneous nucleation are so high 
that liquid water may be trapped in a metastable state up to temperatures much higher than the normal boiling temperature $375\,$K~\cite{Vogel2005}. 
In this respect, vapor generation at the nanoscale shares similarities with the process of explosive boiling characterizing homogeneous nucleation~\cite{Sheperd1982,Glod2002,Zhao2000}. 
Both phenomena are of explosive nature as a result of the instability nature of the phase transition, 
in sharp contrast with the usual nucleation scenario~\cite{Carey,Lowen1992}. 
Yet, these situations open new questions in liquid state physics at small scales~\cite{Lombard2014,Hou2015,Wang2018Jul,Wang2017ACS,Jollans2019}, 
where phase changes are governed by nanoscale effects 
including Kapitza resistance and ballistic vapor transport~\cite{Lombard2016}. In particular, nanobubble threshold has been shown to be sensitive, among others, to the nanoparticle size~\cite{Metwally2015,Lombard2017,Wang2019}, shape~\cite{Fales2019}, surface coverage~\cite{Ogunyankin2018}, laser pulse width~\cite{Wang2018opt} and to the presence of dissolved gas~\cite{Wang2018Jul}.


%
It has been long recognized that 
nanobubble generation around optically heated nano and micro-objects~\cite{Lin1998} 
induces the emission of strong pressure waves~\cite{Sheperd1982,Zhao2000,Glod2002}. 
One macroscopic consequence is a giant photomechanical effect~\cite{Kavokine2020}, 
seen in interacting assemblies of nanoparticles with negative thermophoresis.   
At the microscopic scale, 
the acoustic emission concomitant to nanobubble generation 
opens the way to destruct cancer cells at a subcellular level~\cite{Volkov2007,Brujan2017}. 
However, the very physical cause of cell membrane poration and eventual cell death remains debated.
There are indeed two mechanisms according to which vapor nanobubbles may eventually cause cell destruction.  
In a first scenario, the acoustic pressure waves produced by water vaporization 
alter the cell membrane integrity~\cite{Pustovalov2008}. 
In a second scenario, the expansion of plasmonic nanobubbles 
generates compressive strains of a few percents causing cell membrane breaching~\cite{Zharov2005,Zharov2006,Lapotko2006,Wen2009,Brujan2017,Rau2006,Yao2020}. 
In deciding which scenario actually holds, 
a quantitative understanding of the phenomena  induced by heated nanoparticle is essential.  
In contrast to  nanobubble-based cell therapy where  mechanical damage is to be maximized, 
photoacoustic imaging seeks to minimize acoustic emission 
in its quest toward non-invasive and non-destructive technique~\cite{Ju2012}. 
This is only possible with a fundamental understanding of acoustic emission 
subsequent to explosive vapor formation around illuminated nano-objects. 

Despite their importance for applications of photoacoustic conversion, 
the experimental investigation of nanobubble-induced pressure waves has remained  scarce in the literature~\cite{Sarimollaoglu2014,Wang2018b}. 
On the computational side, 
a body of work dealt with the modeling of photoacoustic effects 
under moderately intense laser excitation~\cite{Calasso2000,Prost2015,Grosges2016,Cavigli2020}.
Relatively little attention has been paid to the case of photoacoustic effects induced by nanocavitation, 
with noticeable exceptions~\cite{Sun2000,Volkov2007,Gonzalez2010,Brujan2017,Wen2017,Shi2017,Gao2020} 
or in the context of laser irradiated retina~\cite{Faraggi2005}. 
However, 
most of these modeling studies rely on two simplifying assumptions. 
The first is on the thermodynamic state of water following quick energy transfer from the particle: 
 saturation is assumed for the liquid  all  the way to its vaporization. 
The second assumption is on the nature of acoustic emission by nanobubble, 
which is often modeled as a radiation effect.
Whether those assumptions are warranted has remained untested so far. 
%

In this study, we propose for nanobubble-induced acoustic emission  
a well-grounded approach that in contrast with previous descriptions,   
does not rely on simplifying assumptions. 
We use a  hydrodynamic-phase field model to predict bubble formation and pressure wave propagation around gold nanoparticles illuminated at
a wavelength $400\,$nm
close to their surface plasmon resonance and characterize the pressure wave field generated by explosive vapor formation. 
Our  simulations  demonstrate that acoustic emission is a strongly non-equilibrium effect, 
resulting from the quick formation of a vapor/liquid interface 
under the very high temperature gradients generated at the nanoparticle/water interface. 
The result is not only a high  peak in pressure, but also a short rising time to reach it, 
leading to a large growth rate in  pressure. 
These effects are especially pronounced for short laser pulses, 
but long pulses  may  produce echoes of equal amplitude acoustic waves. 
The assumption that 
 locally the thermodynamic state of water follows its saturation line and 
 that acoustic emission is caused by bubble expansion
can not be retained in general, 
as they could actually severely underestimate the phenomena at play. 
As we illustrate by discussing acoustic-mediated cell membrane poration, 
the strong pressure waves irradiated by heated nanoobjects 
have potentially powerful effects. 
Our findings should be helpful to optimize the practical applications 
of plasmonic vapor nanobubble to cancer cell therapy.

The remainder of this article is organized as follows. 
We present in Sec.~\ref{sec:model} our model for the nanoparticle and the fluid that surrounds it. 
Section~\ref{sec:results} details our results on 
the generation and propagation of waves, 
the thermodynamics of the fluid
and the efficiency of the process. 
We discuss in Sec.~\ref{sec:application} an application 
to the mechanism of thermomechanical damage in membrane cells. 
A summary and some perspectives are given in Sec.~\ref{sec:conclusion}

\section{Model}
\label{sec:model}

The physical model has been presented elsewhere~\cite{Lombard2015,Lombard2017}. The reader interested in the results may skip this section and move directly to Sec.~\ref{sec:results}. Generally speaking, the approach is based on a diffuse interface model of the fluid accounting for phase transition and acoustic wave propagation. This type of model has been already proposed and discussed in the literature~\cite{Onuki2005,Lombard2015,Movahedinejad2019} and applied successfully to the shock formation following vapor nanobubble collapse~\cite{Magaletti2015,Magaletti2016}. Here, we consider acoustic emission following vapor nanobubble generation by heated nanoparticle.

\subsection{Nanoparticle}

\begin{figure}[t!]
\begin{center}
  \includegraphics[width=8.5cm]{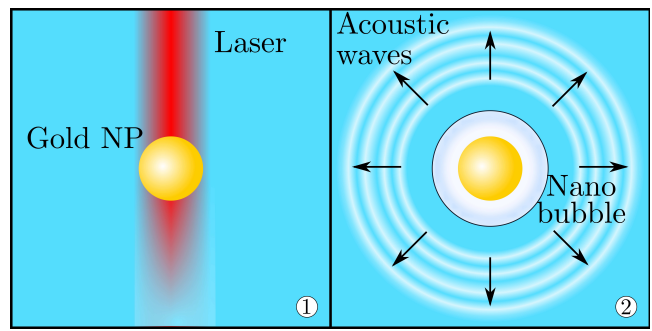}
\end{center}
\vspace*{-0.3cm}

\caption{Illustration of the situation considered: 
a gold nanoparticle in water is quickly heated  by a laser pulse (\ding{172}), 
generating a transient vapor nanobubble  and the concomitant emission of pressure waves that propagate in the liquid (\ding{173}).} 
\label{Fig:illustration_pressure_wave}
\end{figure}

The situation we aimed at modeling here,  depicted in Fig.~\ref{Fig:illustration_pressure_wave}, 
is  a gold nanoparticle immersed in water and illuminated by a laser pulse
with a wavelength $400\,$nm, close to the surface plasmon resonance of the nanoparticle. 
As a result of the strong absorption, 
the temperature of gold electrons $T_e$ rises, 
which induces lattice vibrations through electron-phonon coupling, 
with a  relaxation time $\tauep \simeq 7\,$ps.  
Since we will consider laser pulses with a duration~$\taup$ of at least $10\,$ps,  
we can discard any electron-phonon disequilibrium inside the nanoparticle 
and assign to it a single temperature describing both electronic and vibrational degrees of freedom.    
Besides, 
the temperature within the nanoparticle is assumed to be uniform, 
a reasonable hypothesis owing to the large conductivity of the metal compared to that of liquid water.

In those conditions, the temporal evolution of the nanoparticle temperature $\Tnp$ is  described by 
\begin{eqnarray}
\Vnp \cnp \frac{d \Tnp}{d t} &=& F \signp \frac{\Pi(t/\taup)}{\taup}  - \Snp  \phi, \label{eq:nanoparticle_temperature}  \\
\phi &=& G(\Tnp-\Ts).                                                              \label{eq:conductive_flux} 
\end{eqnarray}
$\Snp$ and $\Vnp$ are the surface and volume of the nanoparticle respectively 
and $\cnp$=2500 kJ/m$^{3}\,$K is the gold specific heat. 
The interaction between laser and nanoparticle  
is described by the size-dependent absorption cross-section \cm{$\signp(\lambda)$} 
as given by Mie theory~\cite{Bohren1983}. 
The laser pulse has fluence~$F$, duration $\taup$ 
and the gate function $\Pi(t)$ is unity for $0<t<1$ and zero otherwise 


The last term in Eq.~\eqref{eq:nanoparticle_temperature} 
involves the flux $\phi$ of heat  leaving the nanoparticle. 
In what follows, we  consider  that this flux 
is proportional to the difference between the nanoparticle temperature 
and the temperature $\Ts$ of the fluid at the surface~\cite{Barrat2003,Merabia2009,Merabia2009b}, 
as prescribed by   Eq.~\eqref{eq:conductive_flux}.  
%
The prefactor is the thermal boundary conductance $G$. 
It keeps a fixed value $G=90$ MW/m$^2$/K, corresponding to a neutral interface (the equilibrium contact angle is around $90$ degrees).
The practical implementation of this contact angle is fully described in Ref.~\cite{Lombard2017}.  

\subsection{Governing equations for the fluid}

To model the fluid dynamics, we resort to a hydrodynamic phase field model based on a free energy density, 
which solves continuum equations for the fluid dynamics based on a diffuse-interface model~\cite{Anderson1998}, 
and which has been successfully applied to address interfacial heat transport and boiling at nanoscale~\cite{Onuki2005,Onuki2007}.
The hydrodynamic conservation equations to describe the dynamics of the fluid around the nanoparticle are 
\begin{align}
\frac{\partial \rho}{\partial t}+  \nabla  \cdot ( \rho \vv ) =&\, 0,  \label{eq:conservation_mass} \\
m \rho(\vcr)  \left( \frac{\partial  \vv}{\partial t}+ \vv \cdot  \nabla  \vv \right) =& - \nabla \cdot \left( \textbf{P} - \textbf{D} \right), \label{eq:conservation_momentum} \\
m \rho(\vcr) c_v(\vcr) \left(\frac{\partial T}{\partial t}+ \vv \cdot  \nabla T \right)=&-l(\vcr)  \nabla \cdot \vv   \label{eq:conservation_energy}   \\ 
& + \nabla \cdot \cm{(k(\vcr) \nabla T)}  +  \textbf{D} :  {\nabla v},   \nonumber
\end{align}
where $\rho$, $\vv$, $T$ stand, respectively, for the number density, the velocity field and the temperature field of the fluid ; 
$m$ is the mass of a fluid molecule, 
$c_v$  and \cm{$k$}  are the fluid specific heat and  thermal conductivity, 
$l = T \left( \partial P /\partial T \right)_\rho$ 
is the Clapeyron coefficient. 
Because these quantities depend on the local thermodynamic state, 
we have emphasized their spatial dependence on the position $\vcr$ in the fluid.  
Finally, 
$\textbf{D}$ and $\textbf{P}$ stand respectively for the dissipative stress tensor and pressure tensor. 
The symbol "$:$" indicates a double dot product. 

These equations are quite general, and may be derived starting from the conservation of mass, linear momentum and internal energy. The pressure tensor and the transport coefficients, however, depend on the specific model retained to describe the fluid. 
In the general heterogeneous and non isothermal situation that we aim at describing, 
the pressure tensor appearing in Eq.~\eqref{eq:conservation_momentum} is not necessarily isotropic, 
and its general expression is given by the so-called Korteweg expression 
\begin{align}
P_{\alpha\beta} = \left[ P_{\rm bulk} -w \rho \triangle \rho - \frac{w}{2} (\nabla \rho)^2  \right] \delta_{\alpha \beta} + w \partial_{\alpha} \rho \partial_{\beta} \rho,  
\label{eq:Korteweg_pressure}
\end{align}
where  the bulk thermodynamic pressure $P_{\rm bulk}$ is  
\begin{equation}
\label{eq:thermodynamic_pressure}
P_{\rm bulk} =-\left(\frac{\partial F}{\partial V}\right)_{\!T}= \rho \left(\frac{\partial f_{\rm bulk}}{\partial \rho}\right)_{\!T}-f_{\rm bulk}. 
\end{equation}
Hence, the pressure tensor contains all the information regarding the local thermodynamics of the fluid and the capillary terms. In particular, the coefficient $w$ penalizes the existence of density gradients $\partial_{\alpha} \rho$, and is related to the fluid surface tension~\cite{Rowlinson2002,Lombard2017}. 

To describe the fluid thermodynamics, we use a van der Waals fluid whose 
free energy density and pressure are 
\begin{align}
f_{\rm bulk}(\rho,T) &= \rho k_B T \left[ \ln \left( \frac{\rho \Lambda ^3}{1-\rho b} \right) -1 \right] - a\rho ^2, \label{eq:f_VanderWaals} \\
P_{\rm bulk}         &= \frac{\rho k_B T}{1- \rho b} -a \rho ^2.                                                     \label{eq:P_VanderWaals}
\end{align}
The van der Waals parameters $a$ and $b$ and the De Broglie wavelength  $\Lambda$
are set so as to represent on the one hand,  the density of liquid water at $297\,$K  and atmospheric pressure 
and on the other hand,  the values of the critical pressure and temperature $P_c$  and $T_c$, 
yielding the  following relation
\begin{align}
\label{eq:critical_param}
P_c     &=   \frac{a}{27 b^2} = 22 \ {\rm MPa},  \nonumber\\
\rho _c &=   \frac{1}{3b} = 322 \ {\rm kg/m}^3,      \\
T_c     &=  \frac{8a}{27b k_B }  = 647.3 \ {\rm K},  \nonumber 
\end{align}
where $k_B$ is the Boltzmann constant. 
The parameter $w$ appearing in the square gradient terms  of Eq.~\eqref{eq:Korteweg_pressure}  
is chosen  to match the surface tension $\gamma=0.72\,10^{-3}\,$J$\,$m$^{-2}$ of water at $T=297\,$K,
as detailed in Ref.~\cite{Lombard2017}. 

A word is of order regarding the value of the  Clapeyron coefficient $l$, 
for which we have considered a dependence in temperature in addition to the dependence in density
\begin{equation}
l (\rho,T) = l_{\rm vap}(T) + \frac{ \rho (r) - \rho_{\rm vap}  }{\rho_{\rm liq} - \rho_{\rm vap}} \left( l_{\rm liq}(T) - l_{\rm vap}(T) \right), 
\end{equation}
where $l_{\rm liq}(T)$ and $l_{\rm vap}(T)$ are the temperature-dependent values of the Clapeyron coefficient 
at a given temperature $T$ for the liquid and vapor of density $\rho_{\rm liq}$ and $\rho_{\rm vap}$. 

\begin{table}[t]
\caption{\label{Table_parametres_thermo} 
Thermophysical parameters in the liquid water  and in the vapor at $297\,$K in SI units.
$\hat{\rho}$ is the mass per unit volume and $C_v$ the heat capacity per unit volume
}
\begin{ruledtabular}
\begin{tabular}{ccccccccc}
phase    & $\hat{\rho}$(kg/m$^3$)   & $C_v$(J/kg/K)   & \cm{$k$}(W/m/K) 	&  $\eta$(Pa\,s)  	    & $l$(Pa)  	     \\
\hline
liquid   & 997.10 	    & 4.13$\,10^3$    &  0.606 	&  8.98 $10^{-4}$   & 5.4 $10^{8}$ 	\\
vapor    & 2.22 $10^{-2}$   & 1.44$\,10^3$    &  0.019 	&  9.9 $10^{-6}$    & 6881  	    					\\
\end{tabular}
\end{ruledtabular}
\end{table}

The dissipative stress tensor is
\begin{equation}
{D} _{\alpha \beta} = \eta \left(  \partial _\alpha v_\beta  + \partial _\beta v_\alpha \right) + \left(\mu - \frac{2 \eta}{3}\right) \, \nabla \cdot \vv \, \delta _{\alpha \beta}.
\label{Eq_tensors_dissipative}
\end{equation}
We assume that the shear and bulk viscosities $\eta$ and $\mu$ 
are related by  $\mu \simeq 5 \eta/3 $~\cite{Schweizer2004} as  is the case for hard spheres liquids.
Since the density $\rho$ in Eq.~\eqref{eq:conservation_mass} is a field with large spatial variations,
we need to account for the variations of the thermophysical and transport coefficients with the local density. For simplicity, we choose a linear relationship between these parameters and the density. 
As an example, the {\em local} shear viscosity is given by 
\begin{equation}
\eta (\vcr) = \eta_{\rm vap} + \frac{ \rho (\vcr) - \rho_{\rm vap}  }{\rho_{\rm liq} - \rho_{\rm vap}} \left( \eta_{\rm liq} - \eta_{\rm vap} \right),
\end{equation}
where the subscripts ${\rm vap}$ and ${\rm liq}$ refer to the bulk values at $297\,$K. 
%
The thermophysical and transport coefficients of liquid water and vapor at $297\,$K  and atmospheric pressure 
are summarized in Tab.~\ref{Table_parametres_thermo}. 

The boundary conditions at the interface between the fluid and the nanoparticle, including the continuity of pressure, \cm{the vanishing of the normal velocity and the no-slip condition for the tangential velocity}, 
are implemented following the scheme described in Ref.~\cite{Lombard2017}.
The temperature evolution of the nanoparticle is coupled to that of the fluid through the conductive flux $\phi$
  and fluid boundary conditions. Specifically, one imposes at the nanoparticle surface the condition~:
  \begin{equation}
    \label{eq:consistency_relation}
    \phi= -k \left(\frac{\partial T}{\partial r}\right)_{r=\Rnp}
  \end{equation}
  where $T$ denotes the fluid temperature and the previous equation is estimated at the nanoparticle position $r=R_{\rm np}$.
We have not taken into account the nanoparticle expansion due to its heating, as we have observed that this expansion plays a minor role in the amplitude of the emitted pressure wave and on the nanobubble dynamics.
As detailed in Ref.~\cite{Lombard2015}, 
perfectly match layers are implemented at the simulation domain boundaries to avoid any reflection
of the pressure waves generated by the sudden vapor generation. 

We consider a spherically symmetric system that allows us to focus only on the radial component of the fields. 
The nanoparticles have a radius ranging from $5$ to $100\,$nm. 
The simulation cell, which is centered on the nanoparticle,  
 has a size $L=260\,$nm for $\Rnp  < 50\,$nm and $L=390\,$nm otherwise. 
We use a time step of $0.2\,$fs and a lattice constant of $0.07\,$nm. 
The velocity field is calculated on a staggered grid, 
shifted from the main grid by half a lattice constant~\cite{Lombard2015}. 

\section{Results}
\label{sec:results}

\subsection{Generation and propagation of the pressure wave}


\begin{figure}[t!]
\begin{center}
\includegraphics[width=8cm]{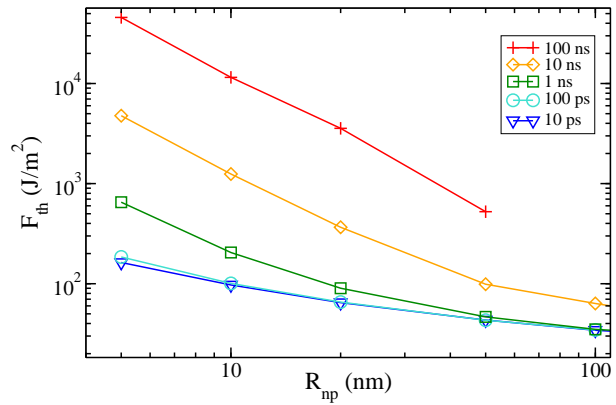}
\end{center}
\vspace*{-0.5cm}

\caption{Vaporization threshold fluence as a function of the nanoparticle radius, for different laser pulse durations.} 
\label{Fig:Fth}
\end{figure}

In this section, we describe the physical processes underlying the appearance and propagation of pressures waves. 
When the nanoparticle is heated by the laser illumination, 
a transient vapor nanobubble is generated, 
provided the fluence is above a threshold value $\Fth$, 
which depends on  nanoparticle size and pulse duration, as shown in Fig.~\ref{Fig:Fth}.  
As demonstrated in~Ref.~\cite{Lombard2017}, 
the  threshold $\Fth$ corresponds to the crossing of liquid/water spinodal temperature, 
$\Tsp \simeq 550\,$K for water, at a distance $1\,$nm from the nanoparticle surface.
In the following, we focus on the  pressure waves that follow vapor nanobubble generation. 

\begin{figure*}[t!]
\begin{center}
\includegraphics[width=7.5cm]{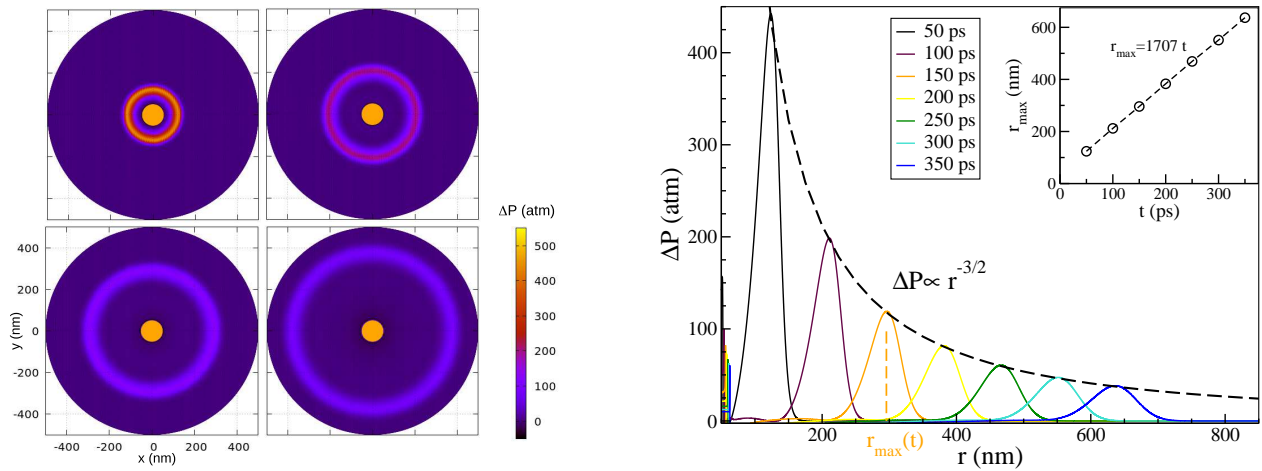} \hspace*{1cm}
\includegraphics[width=8cm]{Fig3b.eps}
\end{center}
\vspace*{-0.5cm}

\caption{
Propagation of nanocavitation-induced pressure waves. 
(Left)
$2$D representation of the pressure wave propagation around the nanoparticle. 
The excess pressure is plotted for different times ranging from $52\,$ps to $252\,$ps. 
The nanoparticle is represented by a yellow circle at the center. 
(Right)
Propagation away from the nanoparticle of the overpressure $\Delta P=P-P_{\infty}$ after the first vaporization event. The dashed line shows the decreasing pressure amplitude while it travels away from the nanoparticle. The inset shows the time evolution of the maximum wave position with time. The corresponding propagation velocity is $c_s=1707\,$m/s. 
The nanoparticle radius is $\Rnp=50\,$nm, the pulse duration $\taup=10\,$ps and the fluence $F=10\,\Fth=430$ J/m$^2$.           
}
\label{Fig:waves}  
\end{figure*}

%
As illustrated in Fig.~\ref{Fig:waves},  
the waves are emitted at the nanoparticle surface and propagate over microns in liquid water.  
The waves are compressive and not tensile in nature.  
Their spatial extension increases during propagation, a dispersion   
that may be qualitatively explained by the density dependence of the compressibility induced by the high pressure levels. 
\cm{Nevertheless, one can infer an effective wave propagation speed, 
whose value $c_w \simeq 1700\,$m/s is found to be slightly above the value of water sound velocity, 
in the initial fluid state prior to any heating. This relatively high value of the apparent sound velocity may be explained 
by two factors~:~first, for the fluence levels considered here, the overpressure corresponding to the wave propagation 
is more than one order of magnitude higher than the initial pressure. Under these conditions, the apparent sound velocity is the one characterizing the high pressure fluid, which is typically $10$ percents higher than the     
initial sound velocity~\cite{Barlow1967}. Secondly, due to the large temperature gradient in the vicinity of the hot nanoparticle,
 the early stage of acoustic propagation does not occur stricly speaking under isentropic conditions. Therefore, the apparent sound velocity may differ   
from its isentropic counterpart}.
As regards the wave amplitude, 
it decays spatially as $r^{-\alpha}$ with an exponent $\alpha = 1.5$ 
that is typical of laser-induced stress wave propagation~\cite{Doukas1996,Vogel1994}. 
that is typical of laser-induced stress wave propagation~\cite{Doukas1996,Vogel1994}. 
we investigate the pressure amplitude at a fixed distance $d$ from the nanoparticle surface.   
We choose $d=500\,$nm,  a typical distance in a biological cell environment. 
Note that in Fig.~\ref{Fig:waves}, 
the fluence level  is fixed to $F=10\,\Fth\simeq 400$ J/m$^2$, 
which is typical of experimental investigation of nanobubble induced cell injury~\cite{Pitsillides2003,Yao2005,Kitz2011,Zharov2006,Zharov2003}.  
In spite of this relatively moderate fluence and even after propagating over $500\,$nm, 
the pressure amplitude level may reach very high values, around  hundreds of atmospheres.

\begin{figure}[ht!]
\begin{center}
\includegraphics[width=7.5cm]{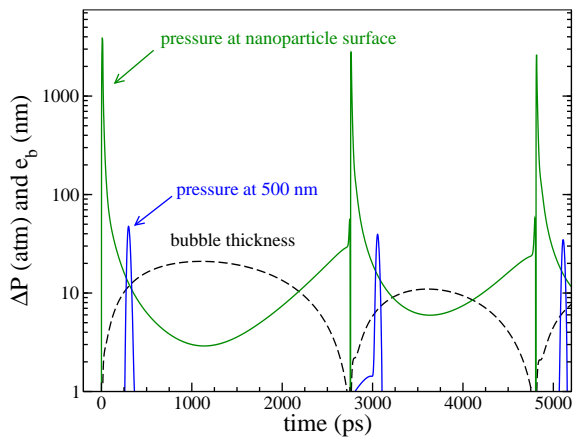} 
\end{center}
\caption{Pressure at the nanoparticle surface and at a distance $d=500\,$nm from the nanoparticle surface, as a function of time. 
The figure also shows the temporal evolution of the bubble thickness $e_b(t)=R_b(t)-\Rnp$. 
The average velocity of the first bubble is $v_b \simeq 16$ m/s. 
The nanoparticle radius is $\Rnp=50\,$nm, the pulse duration $\taup=10\,$ps and the fluence $F=10\,\Fth \simeq 430$ J/m$^2$.}
\label{Fig:formation}
\end{figure}

%
The physical process at the origin of wave formation is scrutinized in Fig.~\ref{Fig:formation}. 
At the nanoparticle surface, a strong pressure increase is observed before the end of the laser pulse ($t=10$ ps). 
The excess pressure starts to relax before vaporization occurs ($t=15$ ps). 
Inspection of the data also indicates that the pressure is maximum just before the bubble forms. 
The next step is a very sharp decrease of the pressure at the nanoparticle surface, 
due to the generation of the vapor phase and the subsequent huge decrease of water temperature after bubble forming. 
This phenomenon has been already observed in Ref.~\cite{Lombard2015} 
and explained by the huge drop of thermal conductance at the nanoparticle/fluid interface following water phase change.

\begin{figure*}[t!]
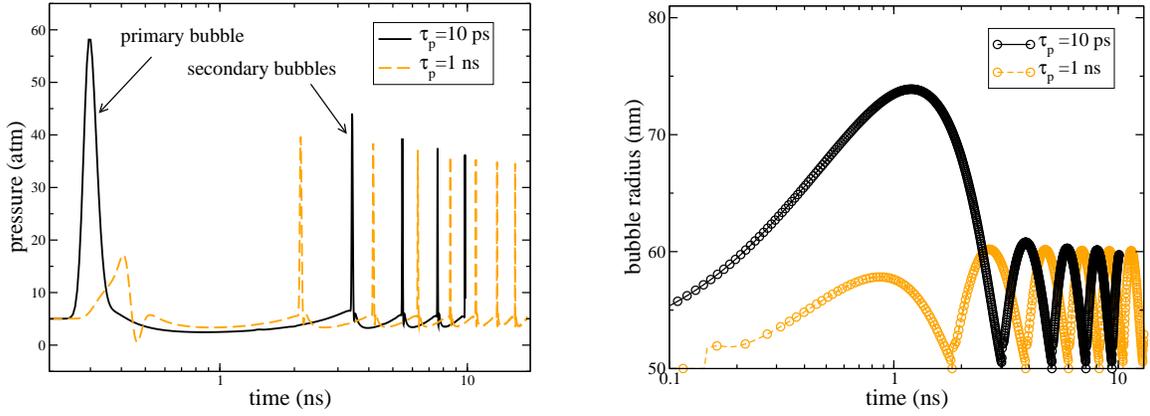

\begin{center}
\includegraphics[width=7cm]{Fig5a.eps} \hspace*{1cm}
\includegraphics[width=7cm]{Fig5b.eps}
\end{center}
\vspace*{-0.5cm}

\caption{(Left) Time evolution of the pressure $P$ at a distance $500\,$nm from the nanoparticle surface for two laser pulse durations, 
$\taup=10\,$ps~(continuous lines) and $\taup=1\,$ns~(dashed lines). 
(Right) corresponding evolution of the nanobubble radius $R(t)$. 
The nanoparticle radius considered here is $\Rnp=50\,$nm and the fluence is $F=16\,\Fth \simeq 690$ J/m$^2$.} 
\label{Fig:Press_10ps_vs_1ns}
\end{figure*}

The first pressure wave  is followed by a series of echoes. 
These secondary waves correspond to the generation of secondary bubbles following bubble collapse, 
as evidenced by Fig.~\ref{Fig:Press_10ps_vs_1ns}. 
After the first bubble collapse, a second pressure wave is emitted. 
When the maximum radius is reached ($t \simeq 1000\,$ps), 
the pressure within the bubble starts to increase again. \cm{ However,
the vapor pressure during the collapse does not reach the levels of pressure  
of the acoustic wave emitted}.
Immediately after the first bubble collapse ($t \simeq 2800\,$ps), 
heat transfer from the nanoparticle to the liquid is reactivated, and a second vaporization event occurs, 
accompanied by the generation of a second pressure wave. A third vaporization event is observed at $t \simeq 4800\,$ps and so on.
This repeated emission of waves, reminiscent of a ``hammer'' effect, 
opens the way to generate repeated thermomechanical damage at a micron distance from the nanoparticle surface.
%
Though the phenomena are similar in nature, 
the generation of waves are influenced by the duration of the pulse. 
The case of a short pulse $\taup=10\,$ps and a longer pulse $\taup=1\,$ns are visible in Fig.~\ref{Fig:Press_10ps_vs_1ns}. 
With the short pulse, 
a first large bubble is inflated which is followed by a series of smaller nanobubbles. 
This contrasts with the case of a long pulse, 
in which the first small size nanobubble  is followed by a series of larger secondary bubbles.
Note that the amplitude of the secondary waves does not vary much with the pulse duration.

A remark is in order on the origin of pressure wave. 
It is important to realize that the bubble wall velocity ($v_b \simeq 16\,$m/s) 
is typically two orders of magnitude lower than the pressure wave velocity $c_w$. 
This clearly indicates that  pressure waves are emitted by the explosive phase change due to fluid heating, 
and  that their propagation is not governed by the bubble dynamics.  
Hence, wave propagation can not be seen as a radiation from the expanding bubble, 
a phenomenon which is shown here to be completely negligible.

Besides, acoustic waves do not originate in the collapse of the vapor bubble. Indeed, as can be seen in Fig.~\ref{Fig:formation}, for a short duration pulse of fluence $10\,\Fth$, the maximal pressure at the particle surface is of the order of $3000$ atm, which is two orders of magnitude larger than the pressure during the bubble collapse (on the order of $30$ atm). This latter value is typical of pressure levels predicted by Rayleigh-Plesset types of models. Indeed, for short pulses, Rayleigh-Plesset models predict a peak pressure of $40$ atm for a nanoparticle radius $R_{\rm nm}=40$ nm and a fluence $F=1000\,F_{\rm th}$~\cite{Brujan2017}\footnote{The difference of nanoparticle radius, i.e $R_{\rm np}=50$ nm vs $R_{\rm np}=40$ nm can not explain the gap between the simulation results and Rayleigh-Plesset predictions, as for short pulse the pressure amplitude is found to scale as $R_{\rm np}^{1.5}$ as can be seen in Fig.~\ref{Fig:Press_500nm_log_log}}. In the Appendix, we illustrate the huge gap between the Rayleigh-Plesset predictions and the phase field results for the maximal emitted pressure. This huge gap may be explained qualitatively by the different physical origin of the peak pressure~:~in Rayleigh-Plesset models, the peak pressure corresponds to the pressure emitted by the bubble collapse. In the phase field simulation the maximal pressure is observed at the formation of a highly out-of-equilibrium liquid-vapor interface. This is consistent with the conclusion that Rayleigh-Plesset types of models tend to severely underpredict the intensity of acoustic emission generated by the sudden formation of a vapor nanobubble. 

Another observation that is essential for the applications 
is that the wave rising time~$\taur$ is found to be short, on the order of $100\,$ps. 
This small value can be ascribed to  the highly out-of-equilibrium nature of pressure emission, 
which is concomitant to sudden nanobubble generation. 
As a consequence of this very fast relaxation, 
the maximal pressure at the particle surface is orders of magnitude larger than the inner bubble pressure. 
Therefore, models based on Rayleigh-Plesset equations would completely fail 
to account for the pressure intensity emitted by the nanobubble~\cite{Brujan2017}. 
For the particular case analyzed here corresponding to a fluence $F=10\,\Fth$, the maximum pressure level at a distance $500\,$nm is approximately $50\,$atm.
A second practical consequence is that 
the high pressure levels $\Pmax$ combined with the short rising times $\taur$  yield  
stress gradients that are high enough to induce significant thermomechanical damage in a biological cell environment, 
as we will analyze in detail below.  

%

\subsection{Influence of pulse duration and nanoparticle size}

\begin{figure*}[ht!]
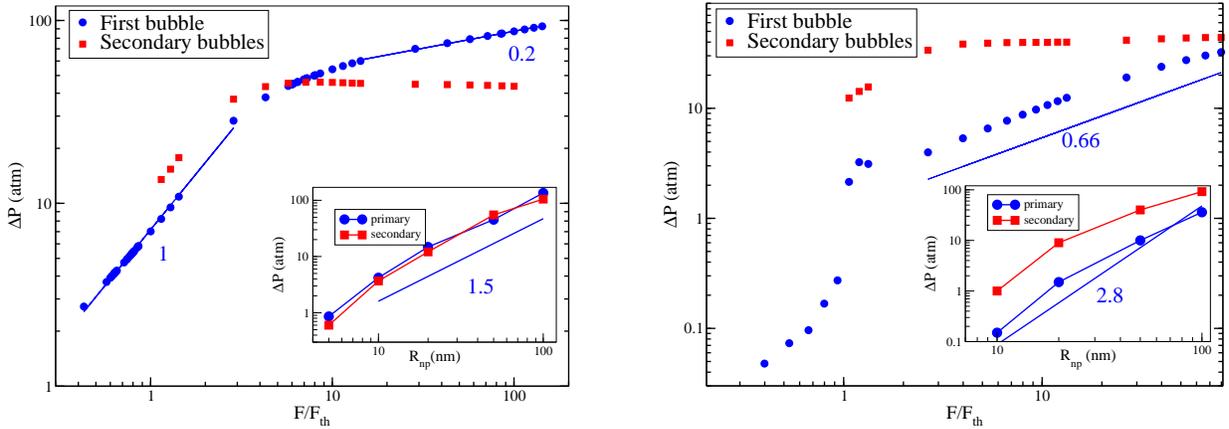

\begin{center}
\includegraphics[width=7.5cm]{Fig6a.eps} \hspace*{1cm}
\includegraphics[width=7.5cm]{Fig6b.eps}
\end{center}
\vspace*{-0.5cm}

\caption{Maximum pressure increase at a distance $500\,$nm from the nanoparticle surface.
The blue circles refer to the first vaporization. 
The red squares correspond to the maximum pressure increase observed among all the secondary vaporizations.
The threshold fluence here is $\Fth=43\,$ J/m$^2$ and the nanoparticle radius $\Rnp=50$ nm.
The insets show the maximum pressure 
as a function of the nanoparticle size~$\Rnp$ for a fixed fluence $F=500\,$ J/m$^2$. 
Two different pulse duration are considered: $\taup=10\,$ps (left)  and $\taup=1\,$ns (right). 
} 
\label{Fig:Press_500nm_log_log}
\end{figure*}

We now investigate how the pressure emission and propagation depend on the laser pulse duration and nanoparticle size. 
The evolution of the pressure wave amplitude, emitted by both primary and secondary nanobubbles, 
as a function of the laser fluence and for different nanoparticle size is quantified in Fig.~\ref{Fig:Press_500nm_log_log}.
Here again, two different behaviours are highlighted, depending on the pulse laser duration $\taup$. 
The strongest effects are observed for the short pulse $\taup=10\,$ps. 
In this case, the pressure amplitude corresponding to the first wave 
increases continuously with the laser fluence when the vaporization threshold is crossed. 
Below the vaporization threshold $F<\Fth$, 
the pressure emitted is due to the fluid dilation induced by local heating. 
The amplitude of the secondary wave increases with the fluence above the threshold $\Fth$ and saturates above a fluence $F \simeq 4 \Fth$.
%
When comparing the amplitudes of the primary and secondary waves, 
they have comparable amplitudes for low fluences but for $F> 6 \Fth$,   
the largest excess pressure are observed for the first vaporization. 
Yet, secondary vaporizations still generate a series of pressure waves of relatively strong amplitude~\footnote{
Remember that, as visible in Fig.~\ref{Fig:Press_10ps_vs_1ns}, 
there are many echoes associated to all the nanobubble rebounds, which have comparable amplitudes.}. 

In the case of long pulse ($\taup=1\,$ns), 
the pressure maximal amplitude levels off when crossing the vaporization threshold. 
At low fluences, $F<\Fth$, the pressure amplitudes are relatively low. 
Beyond $\Fth$, the amplitude is somewhat higher, but still lower than in the picosecond case (for a given value of the fluence). 
In this fluence regime, the pressure amplitude is higher for the secondary vaporizations. 
This implies that cumulative effects may be efficient in the case of long pulse durations. 
Again, the amplitude of the secondary pressure waves saturates for high fluences $F>4 \Fth$.
Note further that the magnitude of the secondary pressure waves is also almost the same for the $1\,$ns and $10\,$ps pulse durations, 
when the fluence is fixed at $F=500\,$J/m$^2$.

The insets of Fig.~\ref{Fig:Press_500nm_log_log} analyze 
the particle size dependence of primary and secondary waves, for a fluence $F=10 \Fth$. 
Again, two behaviours may be distinguished depending on the laser pulse duration. 
When $\taup=10\,$ps, the first vaporization gives the largest pressure, 
while secondary vaporizations are of smaller yet comparable amplitudes.
Overall, the amplitudes of both primary and secondary waves increase with the particle size. 
For $\taup = 1\,$ns, 
the amplitudes of both primary and secondary vaporizations 
increase with the nanoparticle size. 
However, in contrast to the short pulse duration, 
the amplitudes of the secondary waves are stronger than that of the primary waves. 
Again, cumulative effects may be particularly efficient in this regime, as nanobubbles rebounds will give rise to repeated high amplitude pressure waves.

\cm{A remark is in order regarding the possible effect of dissolved gas. All our simulations results concern a situation where both the first bubble and the secondary are made of pure vapor. 
It is legitimate to wonder what would be the influence of dissolved gas. Indeed, as demonstrated experimentally~\cite{Wang2017ACS,Wang2018Jul}, nanobubble generation in the presence of uncondensable gas 
occurs in two steps~:~the first generated nanobubble is made of pure vapor; the secondary bubbles may be enriched by dissolved gas. However, this process occurs after a crossover time corresponding to the diffusion time of the dissolved gas.  
This time is typically in the microsecond range~\cite{Wang2017ACS}. Therefore, we conclude that before this crossover time, all our results obtained in the absence of dissolved gas should hold. Only after the long crossover time, 
which is well beyond our investigated time, the presence of dissolved gas may affect our predictions.}


\subsection{Thermodynamic evolution}

\begin{figure*}[t!]
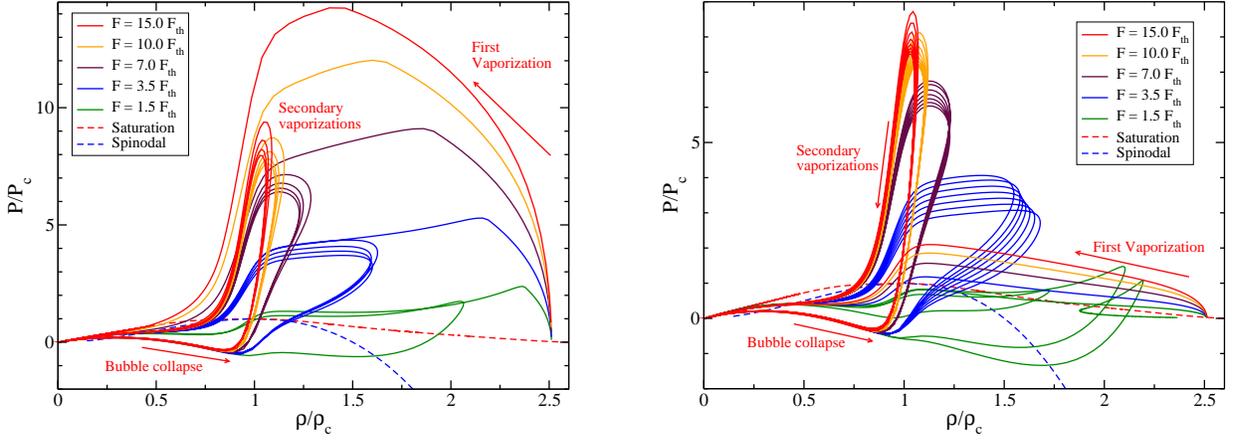

\begin{center}
\includegraphics[width=7.5cm]{Fig7a.eps} \hspace*{1cm}
\includegraphics[width=7.5cm]{Fig7b.eps}
\end{center}
\vspace*{-0.5cm}

\caption{Evolution of the thermodynamic state of the fluid at the particle surface in a pressure-density diagram. 
We consider here $\Rnp=50\,$nm and two different pulse durations:  $\taup=10\,$ps (left) and  $\taup=1\,$ns (right). Here $P_c=217$ atm and $\rho_c=322\,$ kg/m$^3$ are the critical water pressure and mass density.} 
\label{Fig:Phase_Diag}
\end{figure*}

To better interpret the different behaviours observed so far, 
Fig.~\ref{Fig:Phase_Diag} displays the evolution of the thermodynamic state of the fluid at the particle surface 
for two different pulse durations, in a pressure-density ($P-\rho$) diagram. 
The general thermodynamic evolution is similar in all the cases analyzed: 
the liquid prior to any heating is at the saturation value. 
After nanoparticle excitation, the fluid is locally strongly heated on a very short time scale. 
After reaching their maximal values, the pressure and temperature suddenly drop 
before the local density~$\rho$ becomes smaller than the fluid critical density $\rho_c$ and a first nanobubble is generated. 
The nanobubble grows until the local density reaches a minimal density,
after which the bubble collapses. 
As the nanobubble acts as a vapor blanket which thermally insulates the nanoparticle from the fluid, 
the liquid may see a very hot nanoparticle when the bubble collapses,
this latter step occuring under quasi-isothermal conditions, as already shown~\cite{Lombard2015}. 
If the fluence is high enough, a second nanobubble is generated. 
This nanobubble follows another thermodynamic path, 
and Fig.~\ref{Fig:Phase_Diag} shows that its evolution is characterized by loops in the pressure-density phase diagram. 

The thermodynamic path depends strongly on the pulse duration $\taup$. 
For the fastest pulse $\taup=10\,$ps, the fluid pressure reaches a first maximum well above water critical pressure $P_c$ before decreasing slowly until nanobubble generation. 
The maximal pressure levels are always higher for the first vaporization event than for the secondary ones. 
This conclusion is in line with the observation that, for short pulse durations, primary waves yield strongest pressure wave amplitudes. 
By contrast, for the longer pulses analyzed, 
the maximal pressure levels attained before the first vaporization are significantly lower than in the fast pulse case. 
Consequently, for a constant value of the fluence, the pressure emitted after the first vaporization event is considerably lower in the case of long duration pulses.  
The secondary vaporizations generate pressure levels much higher than the first one, which is again consistent with the conclusions reached after analyzing Fig.~\ref{Fig:Press_500nm_log_log}. 

Overall, the analysis of the phase diagram allows us to conclude that in order to produce high amplitude waves, 
the fluid should be brought locally to its supercritical state. 
This may be realized either for the primary bubbles by working with ultrashort pulses, or for secondary waves in all cases, if we rely on cumulative effects created by acoustic echoes. 


\subsection{Analytical model for the pressure amplitude}

In this section, 
we devise a simple analytical model to interpret the evolution of the pressure amplitude, 
as plotted in Fig.~\ref{Fig:Press_500nm_log_log}, as a function of fluence $F$ and nanoparticle size $\Rnp$.
Consistently with the analysis of the phase diagrams displayed in Fig.~\ref{Fig:Phase_Diag}, 
we will consider separately the cases of short and long duration pulses.

\paragraph*{Short pulse.}
When the laser pulse is $\taup=10\,$ps, 
the fluid in the vicinity of the nanoparticle is rapidly heated up to a temperature $\Tmax(F)$. 
Once this temperature is reached, the fluid state follows an adiabatic evolution down to the state point 
where the density matches the critical density $\rho_c$. 
At this moment, the thermal interface conductance drops, the fluid cools down and a vapor layer is generated. 
The maximal pressure, that is the value of the pressure emitted just before nanobubble generation, may be approximated as
\begin{equation} 
    \label{eq:surface_pressure}
    \Delta P_s \simeq  \Pprime (\Tmax-\To), 
  \end{equation}
where $\Pprime \equiv \dPdTrho$ and $\Tmax$ is the maximal temperature at the fluid surface and $\rho_0$ denotes the initial fluid density prior to any heating.
As given in the Supplemental Material (SM)~\cite{refSM}, 
an approximation for $\Tmax$ is available when 
${\it (i)}$ the heating is assumed infinitely fast and 
${\it (ii)}$ the dimensionless number $\cnp \Rnp/\cf l_K$ 
where $\cf$ is the fluid specific heat and $l_K$ is the Kapitza length of the nanoparticle/water interface~\footnote{Denoting as 
$\lamf$} the thermal conductivity of the fluid, the Kapitza length is $l_K=\lamf/3G$.,  
takes values small compared to unity, which is roughly the case here. 
In these conditions,  the maximal temperature increase  $\Delta \Tmax= \Tmax-\To$ is 
\begin{equation}
    \label{eq:maximal_temperature}
    \Delta \Tmax \simeq C_\mathrm{num}  \sqrt{\frac{ G \Rnp}{\cnp \cf \lamf}} \zeta F \sim F \Rnp^{1/2},  
  \end{equation}
Here, $C_\mathrm{num}=\sqrt{3} \dawsonmax/(2 \pi^{3/2}) \simeq 0.084$ is a numerical constant, 
with $\dawsonmax$  the maximum of the Dawson function. 
For notational convenience,  we introduce $\zeta=\signp/\Rnp^3$, 
which for nanoparticles not too large ($\Rnp \leqslant 50\,$nm) is a constant independent of radius. 
The pressure amplitude at a distance~$d$ from the nanoparticle follows 
  \begin{equation}
    \Delta P(d) = \Delta P_s \left(\frac{\Rnp}{\Rnp+d}\right)^{\alpha} \simeq \Delta P_s \left(\frac{\Rnp}{d}\right)^{\alpha}, 
  \end{equation}
  where $\alpha=3/2$ is the exponent describing the spatial propagation of the pressure wave (see Fig.~\ref{Fig:waves}) 
  and we have assumed $\Rnp \ll d$.
Putting everything together and neglecting prefactor, 
the expression of the pressure amplitude at a distance~$d$ is 
   \begin{equation}
    \Delta P(d) \simeq  \Pprime
                       \sqrt{\frac{ G \Rnp}{\cnp \cf \lamf}}  \zeta F \left(\frac{\Rnp}{d}\right)^{3/2} 
                      \! \! \!  \sim   \!  F \Rnp^{2},  
  \end{equation}
The regime where  $\Delta P(d) \sim F$ is clearly visible in Fig.~\ref{Fig:Press_500nm_log_log}.
It will hold until the fluid density in the vicinity of the particle, $\rho_{\rm max}(F)$ approaches the critical fluid density~$\rho_c$.
When this is the case, the thermal conductance at the nanoparticle/water interface 
drops down by orders of magnitude~\cite{Lombard2015}, 
yielding quasi instantaneous fluid cooling. 
This ultrafast temperature drop is accompanied by the emission of a pressure wave, 
and the value of the pressure emitted is close to the fluid pressure at the density $\rho_c$ and temperature $\Tmax$. 
This defines a critical value of the fluence $F_c$ such that $\rho_{\rm min}(F_c)=\rho_c$. 
Beyond $F_c$, the maximal fluid temperature at the surface saturates at a value such that
\begin{equation}
     \Delta \Tmax^c=\frac{\zeta F_c}{\cnp}, 
\end{equation}
where, for $F>F_c$, vapor bubble generation coincides with the crossing of $\Tmax^c$ at the nanoparticle surface. 
In this high fluence regime, the pressure amplitude at a distance $d$ from the nanoparticle surface is
\begin{equation}
    \Delta P(d) =  \Pprime  \Delta T_c \left(\frac{\Rnp}{d}\right)^{3/2} 
                   \sim F^0 \Rnp^{3/2}. 
\end{equation}
This is consistent with the weak dependance of the pressure amplitude observed in Fig.~\ref{Fig:Press_500nm_log_log}, for short pulse $\taup$ and high fluences $F$. 
Furthermore, the scaling $\Delta P \sim \Rnp^{3/2}$ is completely consistent 
with what is observed in the inset of Fig.~\ref{Fig:Press_500nm_log_log}. 
This scaling mirrors the fact that, in this high fluence regime, pressure emission is a local process, 
which is independent of the nanoparticle size.

\begin{figure*}[th]
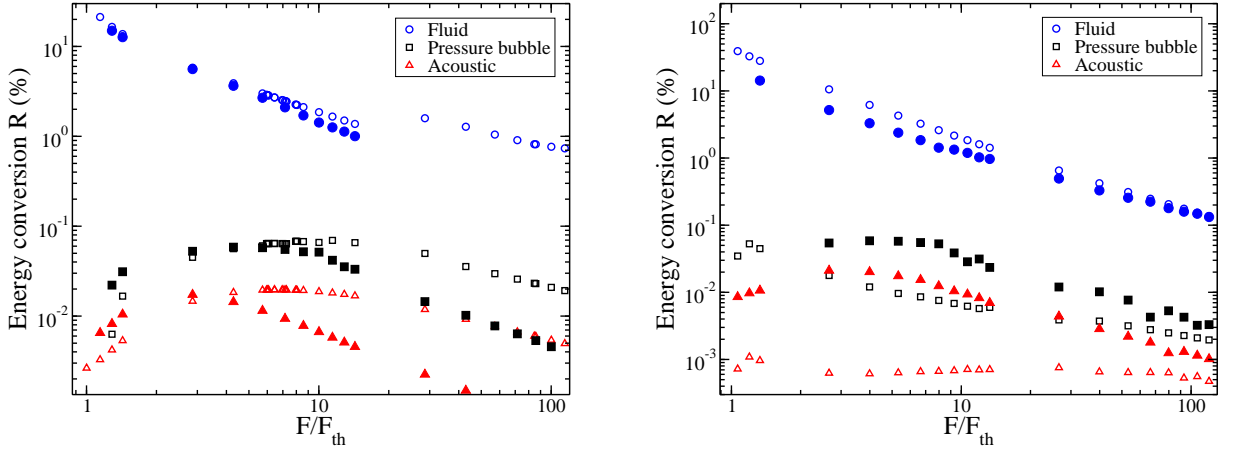

\begin{center}
\includegraphics[width=7.5cm]{Fig8a.eps}\hspace*{1cm}
\includegraphics[width=7.5cm]{Fig8b.eps}
\end{center}
\vspace*{-0.5cm}
\caption{Energy conversion involved in the growth of the bubble and acoustic energy carried by the pressure wave emitted from the nanoparticle surface. Open symbols correspond to primary bubbles-waves while closed symbols are the secondary bubbles-waves.
The pulse durations are  $\taup=10\,$ps (left) and  $\taup=1\,$ns (right).
} 
\label{Fig_Rendement}
\end{figure*} 

\paragraph*{Long pulse.}
Let us now consider the long pulse regime, as illustrated by the $\taup=1\,$ns case in Fig.~\ref{Fig:Press_500nm_log_log}. 
In this latter situation, the pulse duration $\taup$ is longer than the vaporization time~$\tvap$, 
the time necessary for a nanobubble to be generated. 
As a consequence, only a fraction $\tvap/\taup$ of the incoming fluence $F$ serves to vaporize the fluid. 
To estimate the maximal pressure in the vicinity of the nanoparticle, 
we first use energy  conservation to write 
\begin{equation}
  \label{eq:energy_converted}
  \frac{\tvap}{\taup} F \sigma_{\rm np} = \cf \Delta \bar{T}  \frac{4 \pi}{3} \left( (\Rnp+ \sqrt{D \tvap})^3 - \Rnp^3 \right).  
\end{equation}
Here it was assumed that all the energy received serves to heat the liquid in a  spherical shell of thickness $\sqrt{D \tvap}$, 
up to an average temperature $\To+\Delta \bar{T}$. 
The equality is written at time~$\tvap$ where the surface temperature reaches the water spinodal temperature~$\Tsp$. 
Assuming the thickness remains small compared to the particle radius, 
one can obtain  $\Delta \bar{T}$ and deduce a surface pressure 
\begin{equation} 
    \label{eq:surface_pressure}
    \Delta P_s \simeq   \Pprime
                        \frac{\sqrt{\tvap}\, \zeta }{4 \pi \cf \taup \sqrt{D}} F \Rnp . 
  \end{equation}
Now, the vaporization time $\tvap$ may be obtained by considering the short-time behavior of the fluid surface temperature,
as given in the SM~\cite{refSM}: 
\begin{equation}
\tvap = \frac{\pi}{3^{2/3}} \left( \frac{ \cnp^4 \cf\lamf \Delta T_s^2 }{G^4 \zeta^2} \right)^{1/3}  F^{-2/3} \Rnp^{2/3}.   
\label{eq:shortimeapproxTs}
\end{equation}
with $\Delta T_s=\Tsp - \To$. 
One finally obtains the scaling for the pressure amplitude at a distance $d$
\begin{equation}
    \Delta P(d) \simeq \Delta P_s \left(\frac{\Rnp}{d}\right)^{3/2} \sim F^{2/3} \Rnp^{17/6}. 
  \end{equation}
When compared to the numerical data of Fig.~\ref{Fig:Press_500nm_log_log}, 
those scaling laws capture well the main trends.

\subsection{Energy conversion}

Finally, we quantify the energy conversion associated to the acoustic emission. Our reference is the {\em total} energy brought by the laser $\EL =  \signp F$, where $\signp$ is the nanoparticle absorption cross-section and $F$ the laser fluence. Part of this energy is stored in the nanoparticle, and only a fraction of this energy is delivered to the fluid during the bubble growth. We call $\Ef(n)$ the energy which is transmitted to the fluid during the lifetime of the $\nth$ bubble, that is between the collapse of the $(n-1)^{\rm th}$ bubble and the collapse of the $\nth$ bubble.
It is important to realize that not all the energy transferred to the fluid $\Ef(n)$ is used for the growth of the $\nth$ bubble. Indeed, part of this energy is dissipated as heat in water and is therefore not employed for bubble growth. To disentangle this latter energy channel, we have defined $\Esigma(n)$ as the energy effectively converted to mechanical energy. Finally, along with bubble growth, a fraction of the energy transferred to the fluid is converted in acoustic energy, that we denote by $\Ea(n)$. The expressions of $\Ef(n)$, $\Esigma(n)$ and $\Ea(n)$ are given in the appendix. 

Figure~\ref{Fig_Rendement} shows the relative energy conversion channels namely, 
fluid:~$\Ef(n)/\EL$, pressure bubble~$\Esigma(n)/\EL$ and acoustic~$\Ea(n)/\EL$ for the first generated $n=1$ and second bubble $n=2$.
First, we note that most of the energy brought by the laser is stored in the nanoparticle and not transferred to the fluid. 
Second, only a small fraction of this energy is converted in mechanical energy for pressure growth. 
The remaining energy is hence lost in water through thermal diffusion. 
Third, only a tiny fraction of the energy is converted in acoustic energy. 
The magnitude of conversion is found to be consistent to what has been reported for explosive boiling induced by microheaters~\cite{Zhao2000}, and previous estimates of the acoustic energy conversion around photothermal nanoparticles~\cite{Wang2018b}.

The existence of an optimum for the energy conversion (laser energy into mechanical energy) 
may be understood following the same lines as in Refs.~\cite{Lombard2014,Lombard2017}. 
When the laser fluence~$F$ is close to the threshold~$\Fth$, 
the vaporization times are very long. 
In this regime, a significant fraction of the energy transferred to the fluid is dissipated as 
heat, and the energy conversion in mechanical energy is poor. 
By opposition, for large fluences the vaporization time become very short, and in this regime most of the energy supplied by the laser is stored in the nanoparticle. 
This explains why the relative energy to the fluid is a decreasing function of the fluence. 
In between these two regimes, 
there is an optimum where the vaporarization times takes intermediate values, 
for which  the energy transferred to the fluid is not small and the heat losses by thermal diffusion are miminized. 
Close to this optimum in fluence, the relative energy effectively used for bubble growth is maximal.


\section{Application: mechanism of damage to a cell membrane}
\label{sec:application}

\begin{figure}[t]
\begin{center}
\includegraphics[width=8cm]{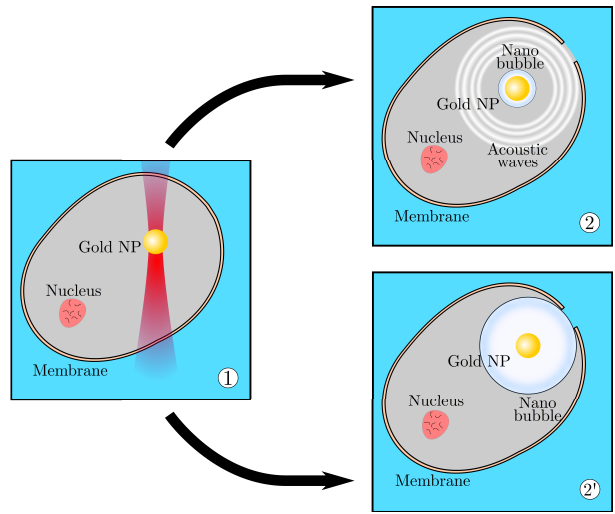}
\end{center}
\vspace*{-0.4cm}

\caption{Illustration of the two possible mechanisms through which nanobubble generation may lead to cell membrane disruption perforation:  
acoustic-mediated perforation (top right) and  
nanobubble expansion and strain induced disruption (bottom right).} 
\label{Fig:Illustration_Mechanisms}
\end{figure}

As a direct application of our results, 
we discuss  here the possible thermomechanical effects induced by nanobubble generation on the destruction or breaching of cell membranes at the subcellular level, as illustrated in Fig.~\ref{Fig:Illustration_Mechanisms}.
Two different mechanisms have been put forward to explain cancer cell destruction following nano vaporization around nanoparticles, namely nanobubble expansion and pressure wave propagation~\cite{Wen2009}.
In the first mechanism, 
nanobubble expansion generates strain inside the cell, and if the cell membrane is deformed by a strain high enough, 
it will be destructed.
For most of the eucaryotic cells, the critical value of the strain is found to be close to
$\epsilon_{\rm cr} \simeq 3\%$~\cite{Boal2002}. 
As regards acoustic emission, 
prior in vitro studies~\cite{Doukas1995,Lee1996} concluded that the control parameter is not the peak pressure but the pressure growth rate 
(or pressure rate of change)~\footnote{Alternatively, 
one could consider the pressure spatial gradient.}:     
\begin{equation}
\sigma= \frac{\Pmax}{\taur},  \label{eq:definitiongrowthrate}
\end{equation}
where $\Pmax$ is the maximum in pressure and $\taur$ the rising time to reach it. 
Typically, a growth rate higher than $\sigma_{\rm cr}=50\,$atm/ns      
leads to cell membrane alteration~\cite{Lee1996}. 
Based on our simulation results, 
we can estimate the possible damage created by illuminated nanoparticle as a function of nanoparticle size, laser fluence and pulse duration.
We concentrate on a fluence interval $F$ in the range $[100, 5000]\,$J/m$^{2}$ $ \simeq [\Fth; 50 \Fth]$, 
corresponding to fluence levels typical of experiments~\cite{Pitsillides2003,Yao2005,Kitz2011,Zharov2006,Zharov2003}. 
To quantify possible damage and decide which of the two above mentionned mechanisms is responsible for membrane disruption, we adopt a simplified model, where the irradiated nanoparticle is at the center of a cell of radius $R_c =500\,$nm.

\begin{figure*}[t!]
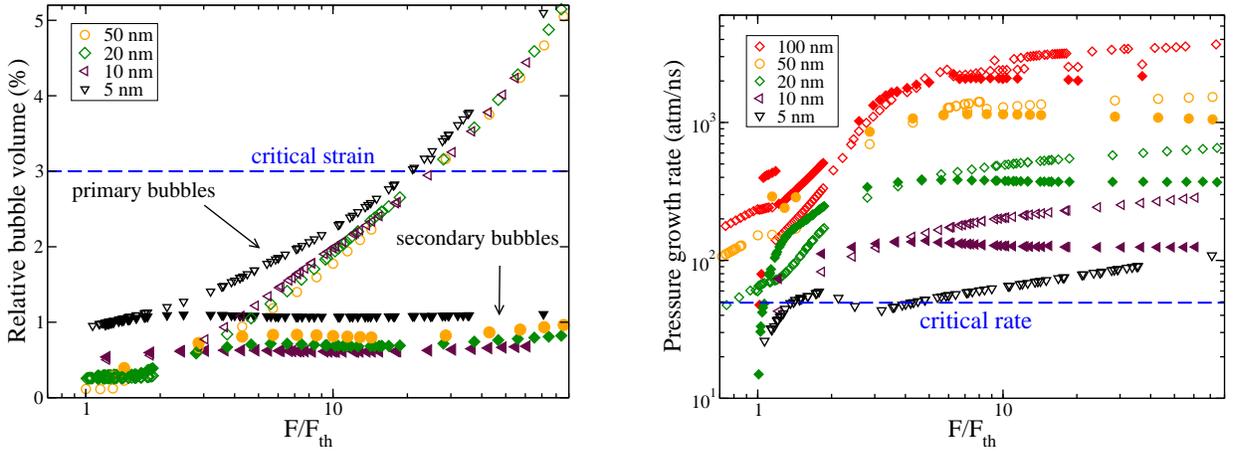

\begin{center}
\includegraphics[width=7.5cm]{Fig10a.eps}   \hspace*{1cm}
\includegraphics[width=7.5cm]{Fig10b.eps} 
\end{center}
\vspace*{-0.5cm}

\caption{
(Left) 
  Maximum relative volume of expanding bubbles, as a function of laser fluence~$F$. 
  Open and closed symbols correspond to the first and secondary nanobubbles respectively.  
(Right) 
  Pressure growth rate, as defined by Eq.~\eqref{eq:definitiongrowthrate}, 
  calculated at a distance $500\,$nm from the nanoparticle surface, as a function of laser fluence $F$ normalized
  by the bubble fluence threshold $\Fth$. The values of the size dependent fluence threshold $\Fth$ are given
    in Fig~\ref{Fig:Fth}.
In both cases, the dashed lines indicate the typical critical value of strain leading to membrane disruption. 
The pulse duration is here $\taup=10\,$ps.  
} \label{Fig:Relative_Volume}
\end{figure*}

Consider first  the effect of nanobubble expansion.  
Given our simplified model, 
the strain is simply equal to the relative bubble expansion 
and its maximal value is $\Delta V_{\rm max}/V= ((e_{\rm max}+\Rnp)^3-\Rnp^3)/\Rnp^3$, 
where $e_{\rm max}$ is the maximal vapor bubble thickness. 
%
%
As visible in  Fig.~\ref{Fig:Relative_Volume}, 
the maximal bubble expansions $\Delta V_{\rm max}/V$  depend only weakly on the nanoparticle size. 
They remain  around one percent for all the secondary generated bubbles, 
while for the primary bubbles, 
they do no exceed two percent  in the fluence range $[\Fth; 10 \Fth]$. 
%
We conclude that membrane destruction due to bubble expansion is quite unlikely below $10\,\Fth$. 
Only with larger fluences $F> 20 \Fth$ would the maximal relative deformation reach the critical value $\epsilon_{\rm cr} \simeq 3\%$. 
In all cases, membrane destruction is only possible for short pulse durations, 
and is triggered by the first bubble generated.


Let us discuss now the effect of the emitted acoustic wave.  
Shown in Figure~\ref{Fig:Relative_Volume} is the growth rate~$\sigma$ as a function of the laser fluence. 
For almost all nanoparticle sizes and fluence above threshold,   
it is above the critical value $\sigma_{\rm cr}=50\,$atm$/$ns.
Therefore, we conclude that the interaction between the pressure wave and the cell membrane 
is the most likely cause of cell damage, at least when the fluence is lower than $1000\,$J/m$^2$.
Nanobubble expansion may cause cell membrane disruption only for higher fluence levels. 

It is thus clear that in our idealized situation, 
acoustic propagation has a more drastic effect than nanobubble expansion. 
This being said, several remarks are in order.
{\it (i)} 
Our simplified model concerned nanobubble generation around a single nanoparticle supposed to be located in the middle of the cell. 
Clearly, the possible damage induced by either mechanism
will depend on the precise location of the nanoparticle, 
and more pronounced effects are to be predicted for nanoparticles in the vicinity of the cell membrane, 
or in the situation where the nanoparticles are inseminated by endocytosis, leading to the mechanical injury of the vesicle. 
{\it (ii)} 
In a real situation, not one but many nanoparticles are inseminated in a diseased cell. 
Given that nanobubble generation is a deterministic process -- contrary to nucleation -- 
the acoustic waves created by the different nanoparticles will interfere coherently 
and the mechanical effect of the acoustic waves will be additive. 
Also, the strain levels seen by the membrane will increase with the number of irradiated nanoparticles. 
We leave an exhaustive study of all these effects to a future study. 
Our discussion around Fig.~\ref{Fig:Relative_Volume} was aimed at balancing the effect of nanobubble expansion and acoustic emission in a idealized situation, and as such we predict relatively low fluence bounds for cell injury. 
In a more realistic situation, collective effects will play a role 
but we anticipate that our conclusions will hold in this case too. 
%
{\it (iii)} 
Finally, we discuss the relevance of another mechanism, as suggested by Volkov et al.~\cite{Volkov2007}. 
The acoustic wave that we characterized has a purely compressive nature 
but it could become tensile upon reflection by organelles floating in the cytosol. 
However, given the typical reflection coefficient of the organelles ($\mathcal{R} \simeq 0.05$ as estimated in Ref.~\cite{Volkov2007}), 
and the pressure peaks of the compressive waves ($\Pmax \simeq 100\,$atm for $F \simeq 5000\,$J/m$^2$), 
this mechanism is quite unlikely to alter the cell integrity.

\section{Conclusion}
\label{sec:conclusion}

To summarize, we investigated numerically the process of pressure emission 
accompanying vapor nanobubble generation around gold nanoparticles illuminated 
by  picosecond or nanosecond laser pulse close to their plasmon resonance frequency.
As the surrounding water  undergoes ultrafast vaporization, 
strong pressure waves are emitted that propagate in the liquid water. 
A recurrent observation is that the processes involved -- in particular the formation of a vapor nanobubble --
are strongly out-of-equilibrium. 

There are two important consequences.  
The first is that the peak pressure emitted from the nanoparticle surface is rather high, 
above the  water critical pressure $P_c=220\,$atm. 
This reflects the spinodal nature of the vapor generation process. 
To maximize the peak pressure, it is clear that the fluid should be brought to its supercritical state, 
which may be achieved either with primary bubbles working with ultrashort pulses, or with secondary bubbles thanks to the nanobubble rebounds.  
The second consequence is that the pressure rising time is short, on the order of $100\,$ps. 
Taken together,  
these two features determine the possible damage that plasmonic nanobubbles may cause in a biological cell environment.  
The combination of high peak pressure and short rising time 
leads to pressure gradient that are sufficiently strong 
to destroy the cell membrane.  
Importantly, this conclusion holds even for fluence levels close to the nanobubble fluence threshold. 
In contrast, 
the relative values of strains generated by nanobubble expansion turn out to be too small 
to induce any noticeable mechanical damage, even for relatively high fluence levels. 
It is only with fluences exceeding the threshold by  several orders of magnitude 
that the strain would become appreciable.

We addressed here an idealized situation where the irradiated nanoparticle is located in a model biological cell.
The possible mechanical damage was assessed based on the damage created by a single acoustic wave. However, it is clear that 
in all situations analyzed, a series of equal amplitude waves should impige the membrane cell. Importantly, the amplitude of these echoes
are only weakly damped as a result of weak viscous dissipation. 
Hence, it should be possible to realize membrane breaching with repeated
action of moderate amplitude nanobubble generated pressure waves. 
The microscopic mechanisms underlying these effects remain partly to be explored. 
A direction for future work is thus to characterize the effects of repeated acoustic waves that emerge from the nanobubble rebounds. 
Combining our modeling approach with molecular dynamics simulations will open the way to a systematic and quantitative investigation of these effects.


%

\begin{acknowledgments}
We are grateful to C.~Loison, T.~Dehoux, A.~Chemin, D.~Amans and F.~Amblard for interesting discussions.   
Part of the computational work herein was carried on in the PSMN Supercomputer at the Centre Blaise Pascal, ENS Lyon (France). 
\end{acknowledgments}

\section*{APPENDIX}

\subsection*{Comparison between phase field simulations and Rayleigh-Plesset predictions}
\begin{figure}[t]
  \begin{center}
\includegraphics[width=8cm]{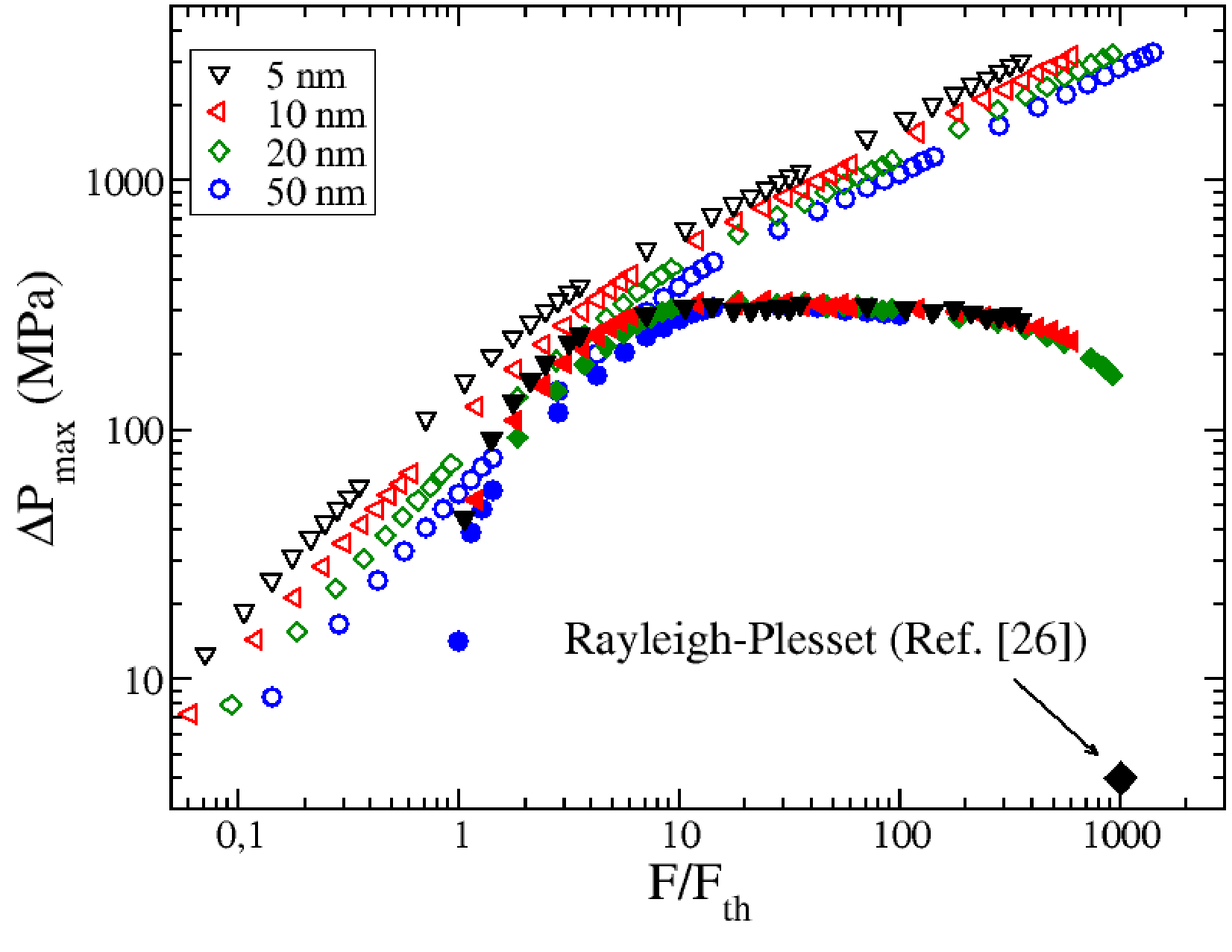}
\includegraphics[width=8cm]{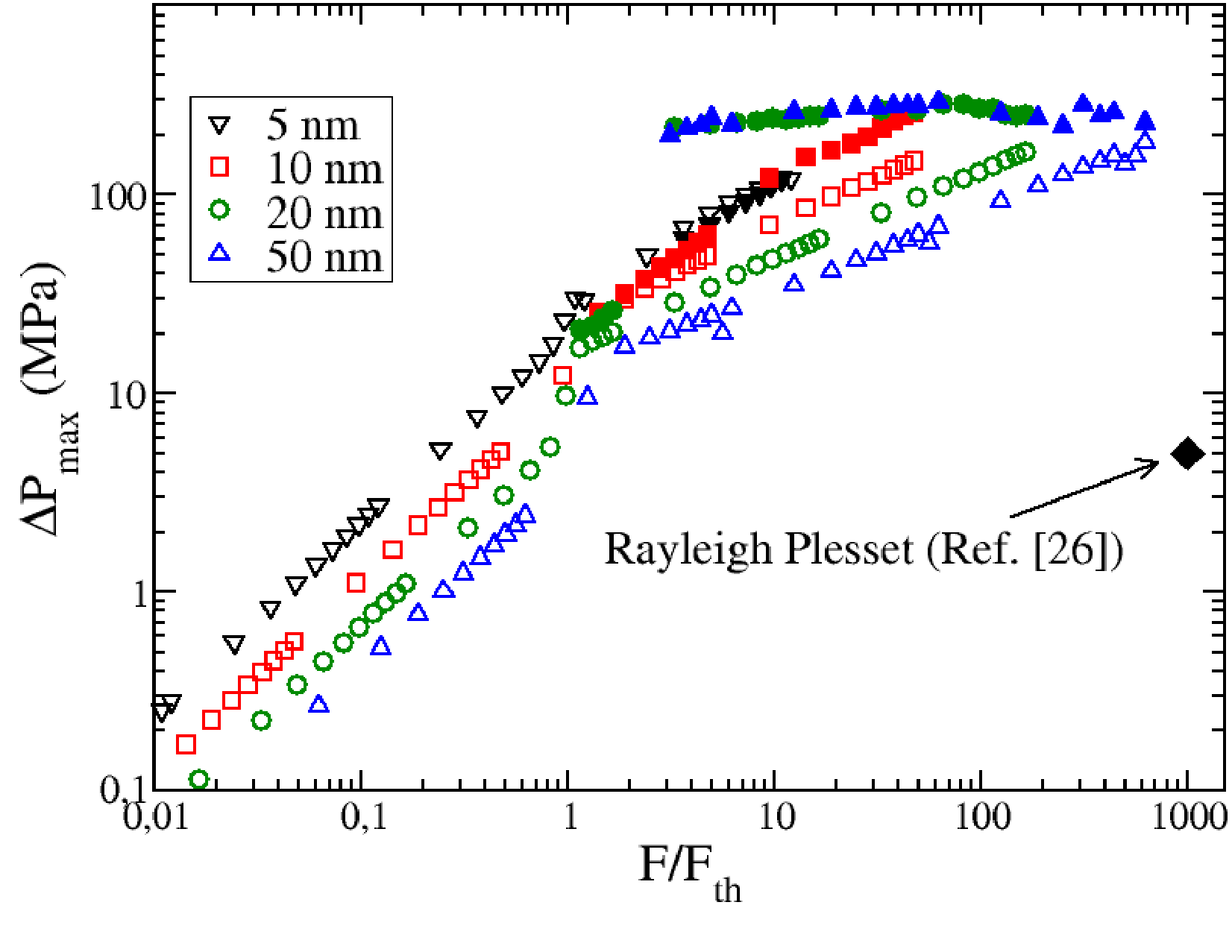}
\end{center}
\vspace*{-0.4cm}
  \caption{Maximal pressure amplitude at the nanoparticle surface as a function of the fluence $F$, and for different nanoparticle sizes~:$\Rnp=5,10,20,50$ nm.
    The values of the size dependent fluence threshold are given in Fig.\ref{Fig:Fth}.
  Top~:~The laser pulse duration is $\taup=10$ ps. Bottom~:~$\taup=10$ ns. Open and closed symbols correspond to the primary and secondary bubbles respectively. 
  For the sake of comparison, we have also represented the Rayleigh-prediction from~\cite{Brujan2017} corresponding to the same pulse durations, and a nanoparticle radius $\Rnp=40$ nm.
\label{Fig:Comparison_simulations_Rayleigh_Plesset}}
\end{figure}

  In this Appendix, we perform a quantitative comparison between the phase field model and the predictions of the Rayleigh-Plesset equation.
  The Rayleigh-Plesset predictions are taken from Ref.~\cite{Brujan2017}. For the sake of comparison, we consider here both a short pulse duration $\taup=10$ ps and
  a long pulse duration $\taup=10$ ns, while the fluence levels range from $F=0.1\Fth$ to $F=1000\,\Fth$. Figure~\ref{Fig:Comparison_simulations_Rayleigh_Plesset} displays the evolution
  of the maximal pressure amplitude as a function of the fluence, for different nanoparticle sizes. As can be seen, the phase field simulations predict
  maximal pressure amplitudes that may be orders of magnitude larger than the Rayleigh-Plesset model. These data suggest that Rayleigh-Plesset types
  of models can not capture the physics of acoustic emission which is essentially an out-of-equilibrium effect.

\subsection*{Energy conversion parameters}
To quantify the relative part of the energy that is transmitted to the fluid, it is useful to introduce the following quantities 
\begin{align}
\deltaTnpnoloss (t) =& \EL/ \Vnp \cnp \times
\left\lbrace
\begin{array}{ll}
1        & \mbox{if\ }  t \geq \taup \\
t/\taup  & \mbox{otherwise} 
\end{array}\right. 
\label{Eq_Tnp_noloss}\\
\Ef(1) =& \Vnp \cnp \left[ \deltaTnpnoloss (t) - \Delta \Tnp ({\rm 1^{\rm st} vap}) \right],  
\label{Eq_Efluid_first_vap} \\
\Ef(n) =&\Ef(n-1) + \Vnp \cnp \left[ \deltaTnpnoloss (t) - \Delta \Tnp ({\rm \nth vap}) \right]  \label{Eq_Efluid_nth_vap} 
  \end{align}
$\EL$ is the energy brought by the laser and 
$\deltaTnpnoloss$  is the increase in nanoparticle temperature 
if all the energy brought by the laser is stored in the nanoparticle. 
%
Equation~\ref{Eq_Efluid_first_vap} is the energy provided to the fluid before the first vaporization (between $t=0$ and $t=t_{\rm vap}$). Equations~\ref{Eq_Efluid_nth_vap} is the energy provided to the fluid for the $\nth$ vaporization, 
that is between the collapse of the $(n-1)^{\rm th}$ bubble and the generation of the $\nth$ bubble. 
%
The energy stored during the growth of the $\nth$ bubble is
\begin{equation}
\Esigma(n)  = 4 \pi \int_{t_{\rm vap}(n)}^{t_{\rm max}(n)} 2 \gamma R_b \dot{R}_b \ dt \label{Eq_energy_Laplace}
\end{equation}
where $t_{\rm max}(n)$ is the time when the $\nth$ bubble reaches its maximal radius, 
$\gamma$ is the fluid surface tension and $R_b(t)$ and $\dot{R}_b(t)$ represent the instantaneous radius and velocity of the $\nth$ bubble. Along with bubble growth, a fraction of the energy flowing to the fluid is converted in acoustic energy. 
The corresponding  expression is
\begin{equation}
\Ea(n) = \int_{\Delta t_{\rm wave}(n)} \int_S  \ea c_s   dS  dt \label{Eq_energy_acoustic} \\
\end{equation}
where $\Delta t_{\rm wave}(n)$ is the time during which the wave generated by nanobubble growth travels through a close surface $S$ at $r= \Rnp$, and $\ea=\Delta P ^2/2 \rho c_s^2$ is the acoustic energy density, with $c_s$ the wave celerity and $\rho$ the fluid density.



\newpage 
\onecolumngrid
\clearpage

\setcounter{equation}{0}  
\renewcommand{\theequation}{S.\arabic{equation}}

\begin{center}
{\bf {\large  Supplementary Material}} 
\end{center}

Here we provide a justification for 
two expressions on the particle surface temperature that were used in the main text. 

For notational convenience, let us introduce two dimensionless parameters $\ga$ and $\beta$ defined from
\begin{equation}
\ga^2 = \frac{\cnp D   }{3 G \Rnp}, \qquad  \beta^2 = \frac{\cnp G \Rnp }{3 \cf \lamf}. 
\end{equation}
We also denote as $\tauk=\cf \Rnp/3G$ the Kapitza time. 
For a nanoparticle with radius $\Rnp=50\,$nm and the parameters considered in the main text, 
$\ga$ is small and $\beta$ typically around unity ($0.16$  and  $1.2$ respectively).

\subsection{Maximum particle surface temperature for low medium heating}

We consider heat diffusion in the medium surrounding a  spherical nanoparticle, 
whose  temperature is assumed uniform. 
Taking $\tauk$ as the unit time and $\Rnp$ as the unit length, the governing equations in dimensionless form are 
\begin{align}
 \partial_t \fnp   =  -  \left( \fnp- \fmeds \right),  \qquad 
 \partial_t \fmed  =  \gamma^2 \; \Delta_r  \fmed ,    \label{eq:sys1}
\end{align}
together with the boundary and initial conditions 
\begin{align}
- \gamma\partial_r \fmed  =  \beta  (\fnp-\fmeds) \mbox{\ \ for\ }r=1, \qquad 
\fnp  =  1,\; \fmed = 0,  \mbox{\ \ for\ \ }  t=0. 
\end{align}
Here, $\fnp(t)$ is the particle temperature, $\fmed(t,r)$ the medium temperature 
and  $\fmeds(t)=\fmed(t,1)$  is the medium temperature at the particule surface. 
$\Delta_r$ is the the radial part of the Laplacian. 
The initial temperature is unity in the particle and zero elsewhere~[S1]. 
Note that because this condition is equivalent to a situation where the particle is impulsively heated at initial time, 
the solution can be seen as a Green function for the problem.

We consider the limiting case $\gamma \rightarrow 0$ with the ratio $\beta/\gamma$ kept fixed.  
Physically, such a situation is reached for a highly conductive medium 
and implies a medium temperature remaining significantly below  the particle temperature ($\fmed,\fmeds \ll \fnp$).  
We solve the equations perturbatively in $\gamma$, focusing on the lowest order only. 
Equation~\eqref{eq:sys1} for the particle temperature immediately leads to $\fnp(t)=e^{-t}$. 
As regards the medium temperature, 
we use a Laplace transform with respect to time ($t \rightarrow s$) to solve the coupled equations. 
The Laplace transform of the surface temperature is found to be $\fmedstilde(s) = \beta \fnptilde(s)/\sqrt{s}$, 
which in time domain yields the relation
\begin{align}
 \fmeds (t) = \beta \int_0^t \frac{\fnp(t-t')}{\sqrt{\pi t'}} dt'.  
\end{align}
Using the solution for $\fnp(t)$ and performing the integration gives
\begin{align}
 \fmeds (t) = \frac{2\beta}{\sqrt{\pi}} F(\sqrt{t}), \label{eq:fsA}
\end{align}
with $F$  the Dawson function.  
Coming back to the variables used  in the main text,
an approximation for the maximal temperature $\Tmax$ reached at the particle surface  is then   
\begin{equation}
 \Tmax - \To = \frac{3 F_\xam^D}{2 \pi^{3/2}} \beta \frac{\zeta F}{\cnp}, 
\end{equation}
where $F_\xam^D \approx 0.541$ is the maximum of the Dawson function. 
This expression is used to derive Eq.~\eqref{eq:maximal_temperature} in the main text.

\subsection{Short-time behavior of surface temperature}
Here we make no assumption on the value of $\gamma$ and $\beta$ parameters. 
Our starting point is the explicit expression for the particle temperature given by Carslaw and Jaeger~[S2]. 
With our notations it reads 
\begin{align}
\fnp(t)   =&  \frac{2 \beta \ga}{\pi} \int_0^\infty 
              \frac{ u^2 e^{- \ga^2 u^2 t }  d u }   
              { \left[ \ga (\ga+\beta) u^2 -1  \right]^2  +\left[ \ga^{2} u^3 -  u  \right]^2}.     \label{eq:T2CJ} 
\end{align}
Now, using Eq.~\eqref{eq:sys1} for $\fnp$ and a change of variable $v=\gamma^2 u^2$, 
one can write for the time derivative of surface temperature 
\begin{align}
\fmeds'(t)   =  \frac{\beta}{\pi} \int_0^\infty e^{-t v} \flap(v) d v, \qquad \quad 
              \flap(v)=\frac{v(v-1)}{\left[ \gamma( \gamma + \beta )v-1 \right]^2 + v(v-1)^2}. 
\end{align}
We considered $\fmeds'(t)$ so that we can directly apply a Tauberian for Laplace transforms. 
Indeed, the integral above is the the Laplace transform of function $\flap(v)$ taken at value~$t$, 
and denoted as $\tilde{\flap}(t)$~[S3]. 
Now, by Tauberian theorem~[S4], 
$\flap(v) \simeq v^{-1/2}$ for $v \rightarrow \infty$. 
implies  
$\tilde{\flap}(t) \simeq \sqrt{\pi} t^{-1/2}$ for $t \rightarrow 0$. 
After time integration, we thus obtain for the short-time behavior of the particle surface temperature 
\begin{align}
 \fmeds(t) \simeq \frac{2 \beta}{\sqrt{\pi}} t^{1/2}.  \label{eq:fsgen} 
\end{align}
Note that because  $F(u) \simeq u$ for small argument, 
the particular case of Eq.~\eqref{eq:fsA} is consistent with the general behavior given by Eq.~\eqref{eq:fsgen}. 
For a heating pulse of unit magnitude starting at $t=0$ and remaining constant afterwards, 
it suffices to integrate the Green function over time,  giving in this case 
\begin{align}
\fmeds(t) \simeq \frac{4 \beta}{3\sqrt{\pi}} t^{3/2}.  
\end{align}
Going back to variables with dimension, 
the short-time behavior of the surface temperature $T(t)$ for a long pulse is at leading order 
\begin{equation}
 T(t) - \To = \frac{\beta}{\pi^{3/2}} \left( \frac{t}{\tauk} \right)^{3/2} \frac{\zeta F}{\cnp}, 
\end{equation}
which is used to derive Eq.~\eqref{eq:shortimeapproxTs}. 

\vspace*{1.5cm}

{\bf References and notes}

[S1] In all cases, only the difference with respect to the initial temperature is considered. 

[S2] Carslaw,~H. \& Jaeger,~J. {\it Conduction of heat in solids}. (Oxford University Press, 1959) 

[S3] Note that here the time~$t$ is the Laplace variable and $v$  is the original variable.  

[S4] Feller,~W. {\it An introduction to probability theory and its applications}. (John Wiley \& Sons, 1968)

\end{document}